\documentclass{JHEP3} 
\usepackage{epsfig}
\epsfclipon

\newcommand{\nco}{\newcommand}
\nco{\beq}{\begin{equation}} \nco{\eeq}{\end{equation}}
\nco{\beqa}{\begin{eqnarray}} \nco{\eeqa}{\end{eqnarray}}
\nco{\lra}{\leftrightarrow}
\def\sfrac#1#2{{\textstyle{#1\over #2}}}

\nco{\sss}{\scriptscriptstyle} \nco{\dphi}{\varphi}
\nco{\lsim}{\mbox{\raisebox{-.6ex}{~$\stackrel{<}{\sim}$~}}}
\nco{\gsim}{\mbox{\raisebox{-.6ex}{~$\stackrel{>}{\sim}$~}}}

\def\pref#1{(\ref{#1})}

\title{\Large Does the small CMB quadrupole moment suggest new physics?}
\author{James M.~Cline$^{1,2}$, Patrick Crotty$^3$, Julien Lesgourgues$^{1,3}$\\
$^1$ Theory Division, CERN CH-1211, Geneva 23, Switzerland\\
$^2$ Physics Department, McGill University,
3600 University Street, Montr\'eal, Qu\'ebec, Canada H3A 2T8\\
$^3$ Laboratoire de Physique Th\'eorique LAPTH, B.P. 110, F-74941
Annecy-le-Vieux Cedex, France\\
E-mail: \email{James.Cline@cern.ch}, \email{crotty@lapp.in2p3.fr}, \email{Julien.Lesgourgues@cern.ch}}

\preprint{McGill-03/09, CERN-TH/2003-099, LAPTH-981/03}

\keywords{Cosmology; Inflation}

\abstract{ Motivated by WMAP's confirmation of an anomalously  low value of the quadrupole
moment of the CMB temperature fluctuations, we investigate the effects on the CMB of cutting
off the primordial power spectrum $P(k)$ at low wave numbers.  This could arise, for
example,  from a break in the inflaton potential, a prior period of matter or radiation
domination, or an oscillating scalar field which couples to the inflaton. We reanalyze the
full WMAP parameter space supplemented by  a low-$k$ cutoff for $P(k)$.  The temperature
correlations by themselves are better fit by a cutoff spectrum, but including the $TE$
temperature-polarization spectrum reduces this preference to a 1.4$\sigma$ effect. 
Inclusion of large scale structure data does not change the conclusion.  If taken seriously,
the low-$k$ cutoff is correlated with optical depth so that reionization occurs even
earlier than indicated by the WMAP analysis.}

\begin{document}

\section{Introduction \label{section:intro}}

One of the intriguing results of the Wilkinson Microwave Anisotropy Probe (WMAP) is a smaller
than expected correlation of temperature fluctuations on large angular scales,
corresponding to a low value for the quadrupole moment. Figure 1 shows the WMAP results
for $C(\theta)$ and the low-multipole $C_l$'s. Their ``Basic results'' paper \cite{basic}
states that the ``probability of so little $C(\theta>60^\circ)$ anisotropy power is $\sim
2\times 10^{-3}$, given the best-fit [running spectral index] $\Lambda$CDM model.'' 
(See also refs.\ \cite{Tegmark,GWMMH} for related discussions.) 
Given the large uncertainties in this region due to cosmic
variance, one might never know whether this constitutes a truly significant deviation from
standard cosmological expectations.  However it has recently been suggested that measuring
polarization from galaxy clusters may in the next decade give a complementary determination of
the quadrupole moment \cite{Bau-Coo} and its evolution with redshift.

\FIGURE{
\centerline{$\!\!\!\!\!\!\!\!\!\!\!\!\!\!\!$\epsfxsize=0.555\textwidth\epsfbox{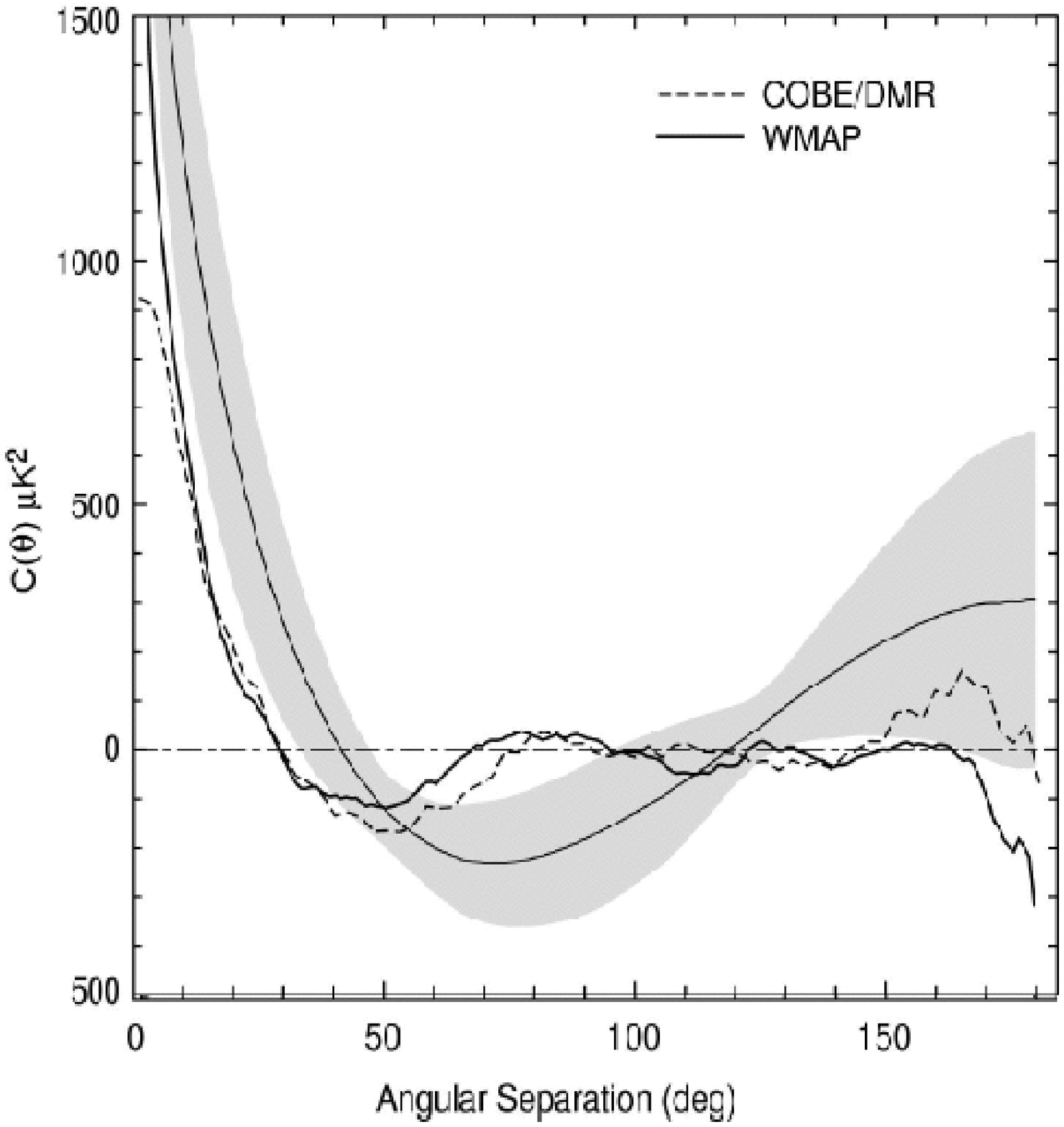}
\epsfxsize=0.5\textwidth\epsfbox{low-l3.eps}}
\caption{\small Left: Angular correlation function $C(\theta)$ for CMB temperature
anisotropy, from fig.\ 13 of \cite{basic}.  Our plot of WMAP TT multipole data 
(from http://lambda.gsfc.nasa.gov/product/map/ map$\_$tt$\_$powspec.cfm)
with the best fit $\Lambda$CDM model (with constant scalar index $n_s$) from table 7 of
\cite{map-params}.  Cosmic variance is a factor of $1\pm 1/\sqrt{l+1/2}$.
}}

Loss of power at large angles could be explained by lowering the primordial power spectrum
of inflaton fluctuations $(P(k))$ at small wave numbers. Prior to the WMAP observation,
there have been several suggestions for generating such an effect. Sharp features
in the inflaton potential \cite{Hodges}-\cite{LPS} can suppress power at low $k$.
Recently \cite{BCLH} studied two other possibilities: an oscillating scalar field $\chi$
coupling to the inflaton can temporariliy suppress inflaton fluctuations, leading to a
cutoff at low $k$; a prior period of matter or radiation domination before the beginning of
inflation has a similar effect \cite{VF}.  Spatial compactness \cite{Uzan} or curvature \cite{Efst}
can also lead to a smaller CMB quadrupole moment. These scenarios require that the total
amount of inflation is close to the minimum amount required to solve the horizon and
flatness problems, or alternatively that the break in the inflaton potential is reached
just when the relevant scales are crossing the horizon.

Recent papers \cite{Bridle,CPKL} considered how cutting off $P(k)$ at low $k$ can
improve the fit to the data.   Ref.\ \cite{CPKL} found that the data favor having such a
modified primordial spectrum at the $2\sigma$ level, whereas \cite{Bridle} obtain weaker
evidence for a cutoff.  The conclusion of ref.\ \cite{CPKL} was based upon using only the
temperature anisotropy data, and ignoring polarization.  The TE cross power spectrum
represents one third of the total data set, and has an important effect on the determination
of parameters that would lower the TT quadrupole moment.  If one simply cuts off $P(k)$ at
low $k$ while keeping all other cosmological parameters fixed, the improvement of the fit to
the TT spectrum is accompanied by a deterioration in the fit to the TE spectrum, such that
the overall improvement of the fit is nil.  To counteract this, one needs to allow other
parameters like the optical depth $\tau$ and the fraction of dark energy $\Omega_{\Lambda}$
to vary.  In the present study, we do a complete likelihood analysis in which all the
relevant cosmological parameters are varied in order to find those values which have the
real maximum likelihood.  We corroborate and extend the results of \cite{Bridle}, who used a
similar approach.

In the second section we describe the theoretical models which can lead to a suppression
of low-$k$ power in the primordial fluctuations.  Section three presents our likelihood
analysis, in which we find that the preference for a cutoff in $P(k)$ is of marginal
statistical significance. Our findings and conclusions are summarized in the final section.

\section{Theoretical Models}

There are many ways to alter the spectrum of inflaton fluctuations relative to the flat
Harrison-Zeldovich form.  Here we will concentrate only on those that produce a sharp
reduction in large wavelength power, needed to suppress the low multipoles of the CMB.  One
way is to engineer the inflaton potential $V(\phi)$, which in principle can yield any
desired form for $P(k)$ \cite{Hodges,CKLL}.   A model which predicts a step-like feature in
$P(k)$  was proposed by Starobinsky \cite{Starobinsky}, which assumes that there is a sudden
change in the slope of $V(\phi)$.  If by chance  the scales presently corresponding  to
large-angle CMB anisotropies exit the Hubble radius at the moment when $\phi$  crosses the
kink in $V(\phi)$, then the power of smaller $k$ fluctuations which subsequently cross the
horizon can be suppressed.  The resulting power spectrum 
(shown in \cite{LPS}) looks similar to the one we
shall discuss below in fig.\ 2b.

More recently (but before WMAP's data release) ref.\ \cite{BCLH} investigated possible
effects of a limited duration of inflation on the power spectrum of the inflaton
fluctuations. Two of these predicted that $P(k)$ should be strongly suppressed at low wave
numbers.
The first was in the context of the hybrid inflation model, with Lagrangian
\beqa \label{eq:modeldef}
    - {\cal L} &=& \sqrt{-g} \Bigl[\sfrac12  \partial_\mu  \phi \,
    \partial^\mu \phi + \sfrac12 \partial_\mu  \chi \, \partial^\mu \chi +
    V(\phi,\chi) \Bigr] , \\
    \hbox{with} \qquad V(\phi,\chi) &=& \sfrac12 \, m^2 \, \phi^2
    + \sfrac14 \, \lambda (\chi^2 - v^2)^2 + \sfrac12 \,
    g \, \chi^2 \phi^2 +  \sfrac{1}{12} \, \tilde\lambda \, \phi^4. \nonumber
\eeqa
It was noticed that if the field $\chi$, which is responsible for triggering the
end of inflation, is oscillating prior to horizon crossing of the relevant inflaton
fluctuations, these oscillations can strongly affect $P(k)$ for $k$ values which are 
below some characteristic scale $k_{\rm max}$, defined by
\beq
	k_{\rm max} = M e^{-Ht_0}
\eeq
Here $M$ is the $\chi$ field mass, and $t_0$ is the amount of time prior to
horizon crossing of the mode $k$ during which $\chi$ was oscillating.  In fact
there are two important scales, this one, which comes about because the $\chi$
oscillations cause production of inflaton fluctuations, and a lower one, below
which the fluctuations are suppressed.  The suppression occurs because the average 
value of $g\chi^2$ contributes to the squared mass of the inflaton,
$\delta m^2_\phi = \langle g
\chi^2\rangle = \frac12 g\chi_0^2 e^{-3Ht}$. 
When $\delta m^2_\phi\gsim H^2$, the inflaton rolls quickly, which suppresses fluctuations
on scales $k$ which have not yet crossed the horizon.  However, the amplitude of $\chi$ 
redshifts exponentially, and at a certain time $\delta m^2_\phi$ will fall below $H^2$.
Fluctuations which have not yet frozen out by this time will have a chance to grow toward their
normal amplitude, $\delta\phi(k)=H/2\pi$; thus larger wave numbers will be unchanged by this effect.
The
fractional amount by which $P(k)$ is reduced at low $k$ is $g\chi_0^2/H^2$ for small
values of this parameter, where $\chi_0$ is the initial amplitude of the $\chi$
oscillations, and $H$ is the Hubble parameter during inflation.  For large values
of $g\chi_0^2/H^2$, the suppression is essentially complete, $P(k)\cong 0$.  This 
is illustrated for some typical parameter values in figure 2a.

The second situation studied in \cite{BCLH} is that in which the inflaton fluctuations with
wave number $k$ cross the horizon shortly after the beginning of inflation, where the
inflationary  epoch is preceded by matter or radiation domination \cite{VF}.  In this case,
power is suppressed on scales below $k \sim H$, the only relevant scale in the problem.  
(This refers to the value of $k$ at the time when inflation starts.  The physical wave
number gets reduced by the subsequent inflation.)  The deformation of the power spectrum has
a unique form, which is shown in figure 2b for the case of prior matter domination.  The
suppression of $P(k)$ in this case is a factor of $0.063$.

\FIGURE{
\centerline{\epsfxsize=0.5\textwidth\epsfbox{newpap-raw.eps}
\epsfxsize=0.51\textwidth\epsfbox{newpap-eta.eps}}
\caption{\small (a) Effect of $\chi$ oscillations on $P(k)$ for 
$M=50H$, $g\chi_0^2/e^{3Ht_0} = 0.003 H^2$, $t_0 = 4/H$.  $P(k)$ is normalized such
that $P(k)=1$ for a flat Harrison-Zeldovich spectrum. (b)
Effect of prior period of matter domination on $P(k)$.  The normalization of $k$ is
in fact arbitrary---see discussion at beginning of section 3.}}

Both of the above effects rely upon having a limited period of inflation: inflation must not
have started much earlier than the time when the large-angle CMB fluctuations first crossed
the inflationary horizon.  For the oscillating scalar scenario, this is because  its
oscillations are quickly Hubble damped.  After at most 30 e-foldings of inflation,
the amplitude of $\chi$ will be too small to have any further effect on the CMB.
In the second scenario, there must be a coincidence of scales such that the large angle
fluctuations cross the horizon just at the beginning of inflation; in this case inflation
can last no longer than the minimum duration needed for solving the horizon and flatness
problems.

\section{Effect on CMB temperature fluctuations\label{section:effect}}

Let us consider the effect of the spectrum shown in fig.\ 2b on the CMB temperature
anisotropy.  The physical scale of wave numbers at which $P(k)$ changes is determined
by the total amount of inflation following the horizon crossing of the corresponding mode,
denoted by $k=H$ in fig.\ 2b.  A longer period of inflation stretches the wavelength of
this mode to larger physical values.  We refer to this model-dependent cutoff scale as
$k_*$.  Figure 3 shows the effect of varying $k_*$ in a modified version of the CMBFAST
code \cite{CMBFAST} which incorporates the distorted spectrum.  There it is seen that
the relevant scales for suppressing the lowest multipoles are on the order of $k_*
\sim$ a few $\times 10^{-4}$ Mpc$^{-1}$.  As noted in \cite{Bridle,CPKL}, even a sharp cutoff
in $P(k)$ does not generate a sharp rise in the temperature
multipoles since the latter are a 
convolution of the former of the form 
$C_l \sim \int dk k^{-1} j_l^2(k[\eta_0 - \eta_{\rm dec}]) P(k)$.

\FIGURE{
\centerline{\epsfxsize=0.75\textwidth\epsfbox{eta3.eps}}
\caption{\small Effect of prior period of matter domination on temperature anisotropy.
The cutoff wave number $k_*$ is in units of Mpc$^{-1}$.}}

If we ignore the TE polarization data, it is possible to obtain a better fit to the
measured WMAP anisotropy in the manner shown in fig.\ 3.  However, if one changes $P(k)$
like this without altering any other parameters, it exacerbates the fit to the TE
spectrum.  Figure 4a shows the experimental data at low $l$ \cite{Kogut}; fig.\ 4b shows
that larger  values of the cutoff wave number $k_*$ suppress the low TE multipoles in
conflict with experimental observation of stronger power at low $l$.  The strong TE 
signal at low $l$ is the basis for WMAP's inference of a large optical depth $\tau$ and
early reionization, which shows that $\tau$ is an important parameter to vary in order
to try to repair the damage to the fit to TE done by modifying $P(k)$.  However changing
$\tau$ can also hurt the agreement between data and theory for the higher multipoles unless
the scalar spectral index $n_s$ is also adjusted to compensate the change in $\tau$.  

\FIGURE{
\centerline{$\!\!\!\!\!\!\!\!\!\!\!$ $\!\!\!\!\!\!\!\!\!\!\!$\epsfxsize=0.60\textwidth\epsfbox{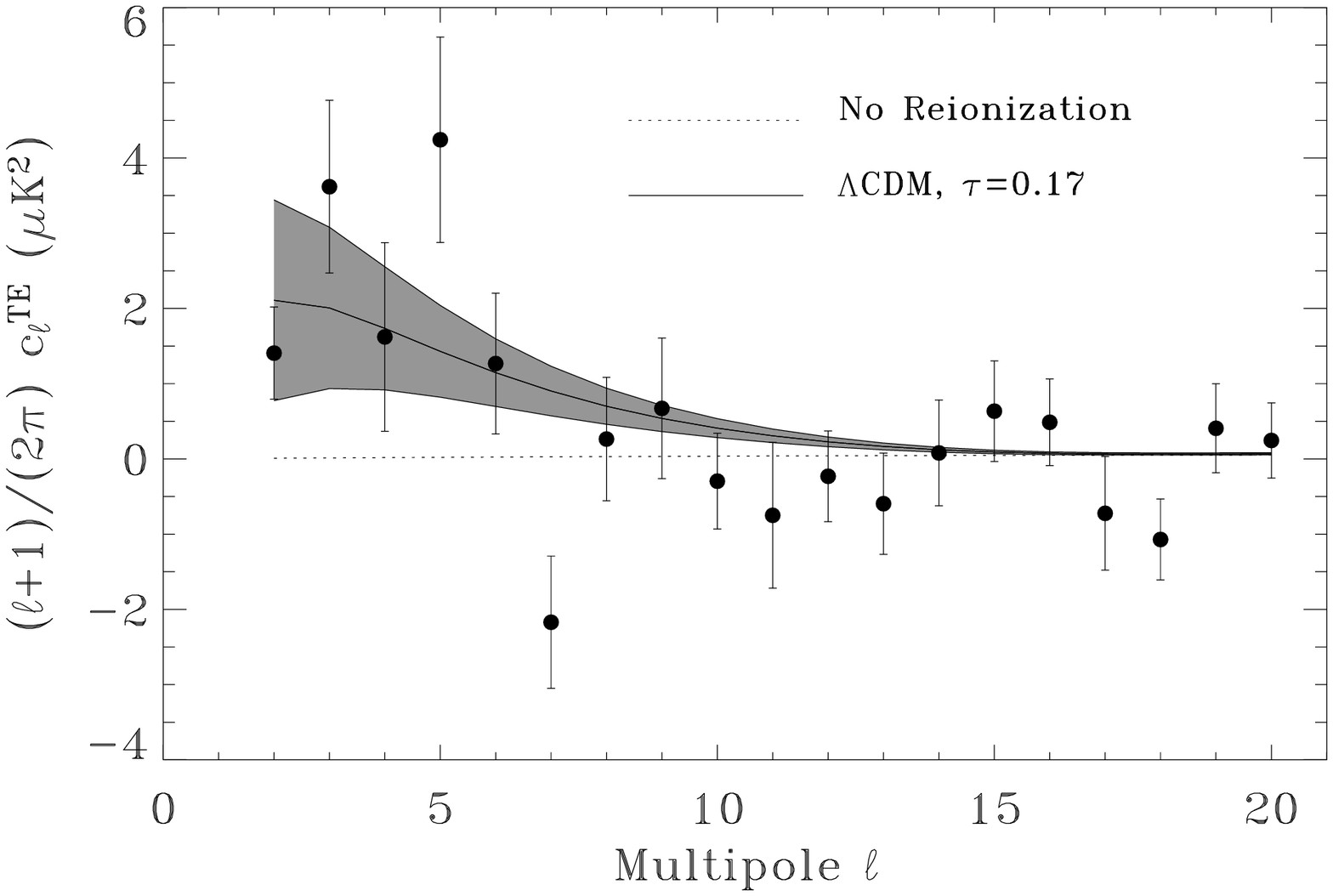}
\epsfxsize=0.52\textwidth\epsfbox{te-sim3.eps}}
\caption{\small (a) WMAP TE cross polarization spectrum (fig.\ 8 of \cite{Kogut}).
(b) Suppression of TE spectrum by $P(k)$ with low-$k$ cutoff $k_*$.}}

This kind of reasoning suggests that no safe conclusions can be drawn without 
doing a
complete analysis in which at least a minimal set of cosmological parameters
(like the six-dimensional parameter space considered in 
\cite{basic,map-params}) are allowed to vary.
We have therefore generated grids\footnote{Our grids contain the points
$\omega_b
\in \{ 0.019,$ $0.022,$ $0.025,$ $0.028 \}$, 
$\omega_{cdm} 
\in \{ 0.08,$ $0.10,$ $0.12,$ $0.15,$ $0.18 \}$, 
$h 
\in \{ 0.5,$ $ 0.6,$ $ 0.7,$ $ 0.8,$ $ 0.9 \}$, 
$\tau 
\in \{ 0.0,$ $ 0.1,$ $ 0.2,$ $ 0.3,$ $ 0.4 \}$, 
$n_s
\in \{ 0.80,$ $ 0.86,$ $ 0.92,$ $ 0.98,$ $ 1.04,$ $ 1.10,$ $ 1.16,$ $ 1.22,$ $ 1.28 \}$
and for the cut-off scale 
$k_*
\in \{ 0,$ $ 10^{-3},$ $ 2 \times 10^{-3}, ..., 10 \times 10^{-3} \}$} of
theoretical power spectra with varying values for
$\omega_b \equiv \Omega_b h^2$, $\omega_m \equiv \Omega_m h^2$, 
$h$, $\tau$, $A_s$, $n_s$ and for
the cutoff scale $k_*$, using several forms for the cutoff spectrum 
$P(k)$. We fixed to zero some parameters which are not strictly necessary
for fitting the WMAP data, like the spatial curvature or the amount of
primordial gravitational waves.  In the 
following, the minimum $\chi^2$ values have been found using a Powell
minimization  algorithm. For models which do not match one of the grid
points, our code first computes the power spectra $C_l^{TT}$,
$C_l^{TE}$, $P(k)$ by cubic interpolation between the grid's neighboring
points, and then calculates the corresponding $\chi^2$ value. We checked
that for our grid spacing, the interpolation is always 
accurate to one per cent.
To be sure that we understand the results of ref.\ \cite{CPKL}, we started 
with their 
ansatz
\beq
\label{ansatz}
	P(k) = A_s(1 - e^{-(k/k_*)^\alpha})k^{n_s-1}
\eeq
where $\alpha=3.35$ is chosen to match the shape at low $k$ of a model similar to  that
shown in fig.\ 2b (they considered a period of prior kination rather than matter
domination).  We performed comparisons with three different data sets: WMAP TT power
spectrum alone, WMAP TT and TE combined spectra, and WMAP TT$+$TE supplemented by large
scale structure data from the 2 degree field (2dF) galaxy redshift survey \cite{2dF}. For
WMAP, we computed the likelihood of a given model using the software provided by the
collaboration and described in \cite{Verde}. For the 2dF power spectrum, we used the window
functions and correlation matrix available at http://msowww.anu.edu.au/2dFGRS/.

First, we verify that there is some preference for a nonvanishing cutoff  in $k$ space
when only the temperature data is included. Figure 5 shows the likelihood (normalized to
1 at the best fit value) for $k_*$, having marginalized over the other parameters. The
model without a cutoff is excluded only at the 90\% confidence level ($\Delta \chi^2
\simeq 2.6$), which is not very significant.   We note that artificially changing the first
three theoretical multipoles, so as to exactly match  the first three data points, 
improves $\chi^2$ by 7.0, corresponding to a Bayesian confidence level of
more than 99\%. Had we been able to achieve such a large $\Delta\chi^2$, we would have
been more convinced that  there is some evidence for a cutoff power spectrum in the
temperature data.  

We obtain a smaller preferred value of $k_*\cong 0.0003$ Mpc$^{-1}$ than did  ref.\
\cite{CPKL}, who found $k_*\cong 0.0005$ Mpc$^{-1}$. We have verified explicitly that
this discrepancy is due to their  having done the analysis on a smaller grid, in a
restricted subspace of the parameters,  which does not include the true minimum
$\chi^2$. Ref.\ \cite{CPKL} only  explored the space of the parameters which have a
direct effect on the low--multipole temperature spectrum,\footnote{The  parameters
$\Omega_{\Lambda}$ and $n_s$ affect both small and large multipoles. So, for
consistency, the temperature power spectrum should be fitted with no restriction on
$\tau$, $\omega_m$ and $\omega_b$, in order to have any possibility of compensating for
the effects of ($\Omega_\Lambda$, $n_s$) at large $l$.} namely the primordial parameters
$A_s$, $n_s$, $k_*$ and the cosmological constant $\Omega_\Lambda$ (which can be
re-expressed in our basis as $\Omega_\Lambda = 1 - \omega_m / h^2$). Other parameters
were kept fixed at $\omega_m =0.135$, $\omega_b = 0.022$,  $\tau=0.17$. These
constraints correspond to the  best-fit values (with a running $n_s$) for the
``WMAPext+2dFGRS''  data set which includes data from other CMB experiments and large
scale structure measurements.   {\it A priori}, there is no justification for imposing
such constraints on a smaller data set  (WMAP TT alone). For instance, we find that in
the absence of a  cutoff ($k_*=0$), the maximum likelihood in the full six-dimensional 
parameter space has an effective $\chi^2$ of 972 for 893 degrees of freedom,  while
imposing the previous constraints raises the best $\chi^2$ to 979. When polarization and
large scale structure data are included in the  analysis, fig.\ 5 shows that the
evidence for a cutoff spectrum gets even weaker, while values of $k_*$ bigger than $5
\times 10^{-4}$ Mpc$^{-1}$  are now excluded at $2\sigma$.

\FIGURE{
\centerline{\includegraphics[angle=270]{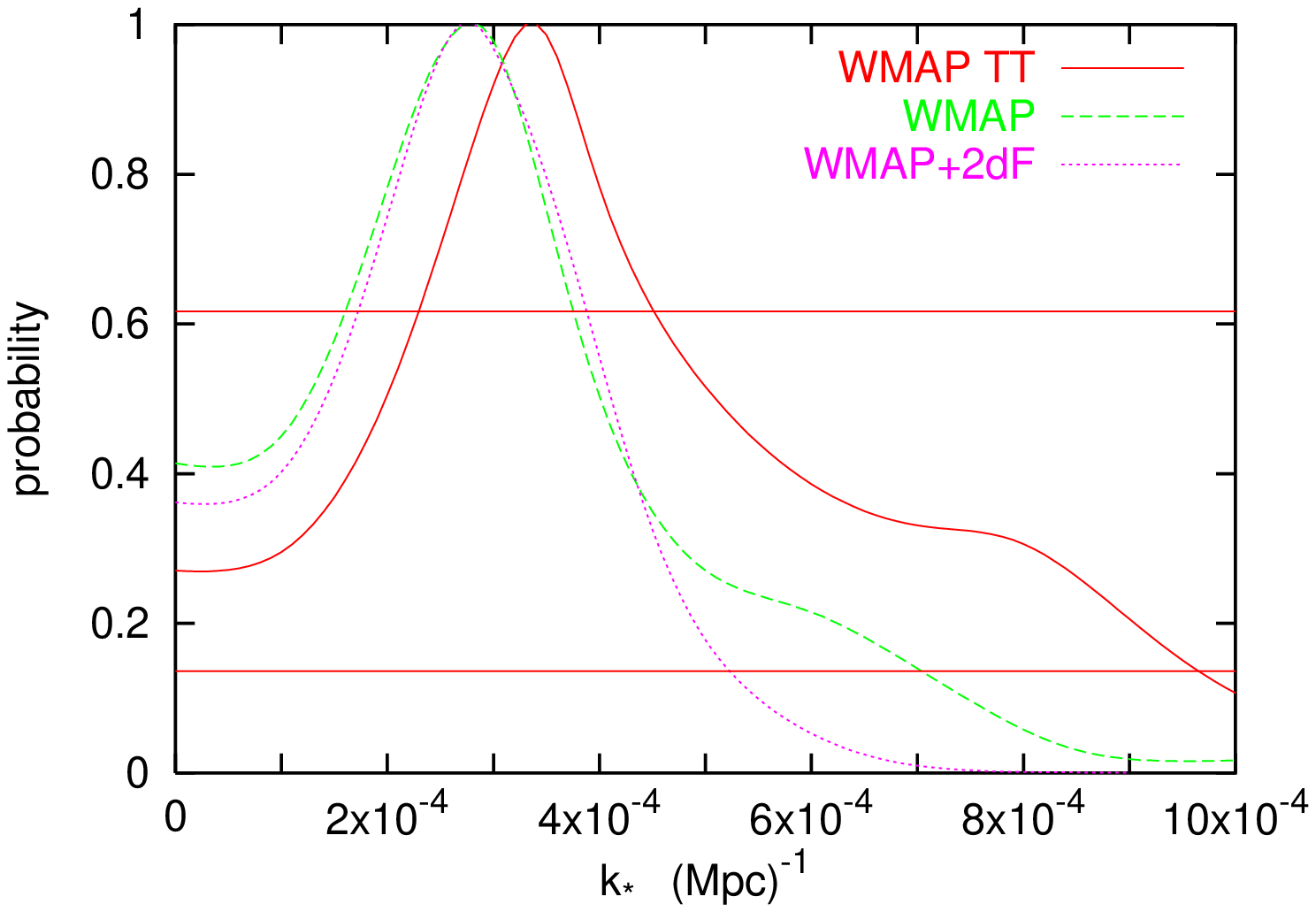}}
\caption{\small Likelihood of the wave number cutoff $k_*$ 
for WMAP TT, TT$+$TE
and TT$+$TE$+$2dF data sets.  
The TT curve has a slightly higher preferred $k_*$
relative to the other two. The horizontal lines refer to the
1$\sigma$ (68 \%) and 2$\sigma$ (95 \%) confidence levels.}}

To see how parameters are correlated, in fig.\ 6 we also show
confidence regions in the $\tau$-$k_*$ plane for the three different
data sets, and the maximum likelihood value of each parameter for fixed values 
$k_*$. Using TT alone (fig.\ 6a), the preference for nonzero
$k_*$ is only slightly bigger than a $1\sigma$ effect. When TE polarization 
data is included (fig.\ 6b), the significance is further reduced, and the
anticipated need for larger optical depth values is manifested.  By 
examining the most
likely values of the other parameters as a function of $k_*$, fig.\
6d, we see that an increase in $k_*$ and $\tau$ 
can be compensated by an increase
in $n_s$, $h$, $\omega_b$, and a decrease
in $\omega_{\rm cdm}$.  Inclusion of large
scale structure data (fig.\ 6c) breaks this degeneracy, but does not
much improve the low statistical significance of the determination of
$k_*$.

\FIGURE{\centerline{\epsfxsize=0.6\textwidth\epsfbox{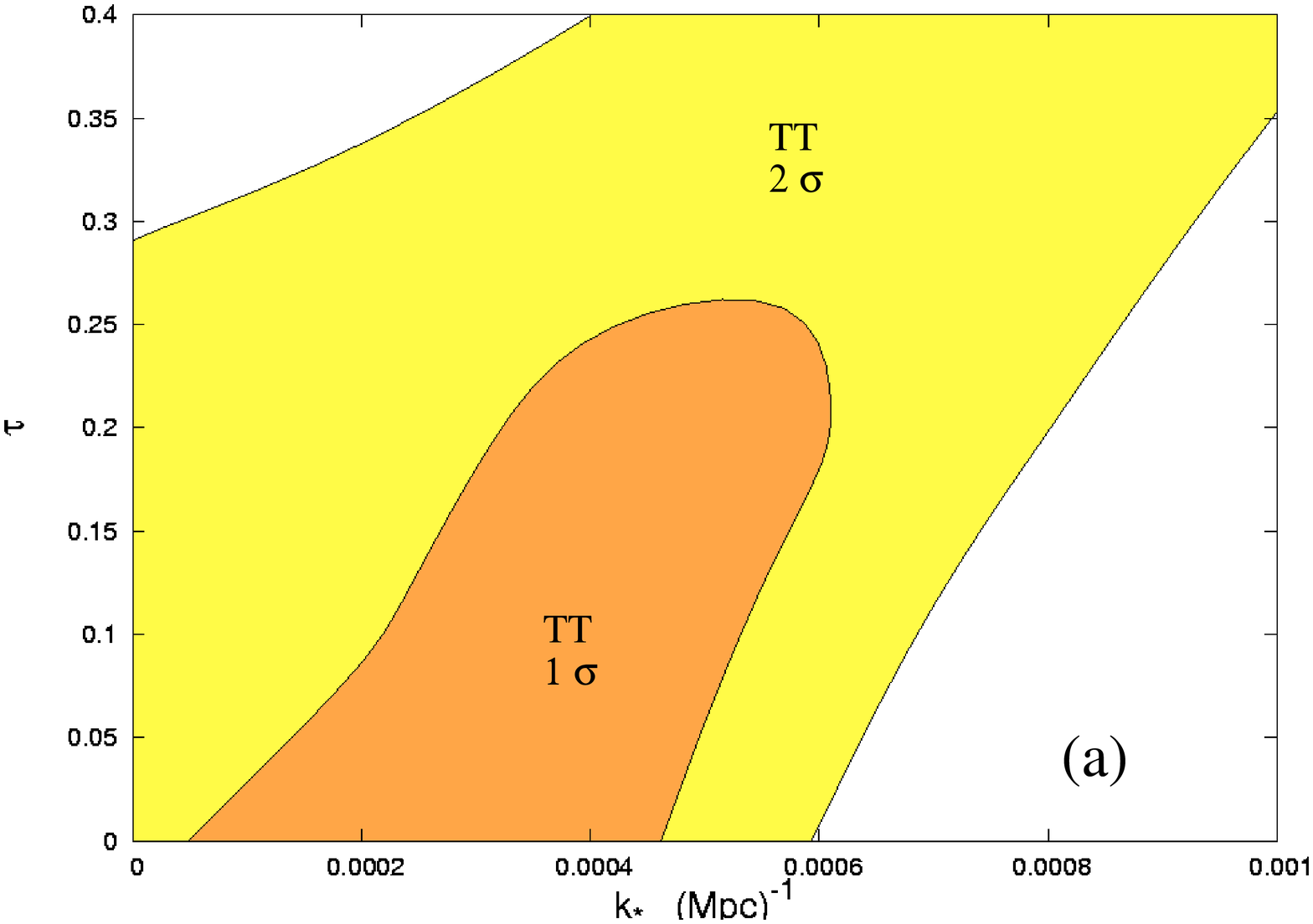}
\epsfxsize=0.6\textwidth\epsfbox{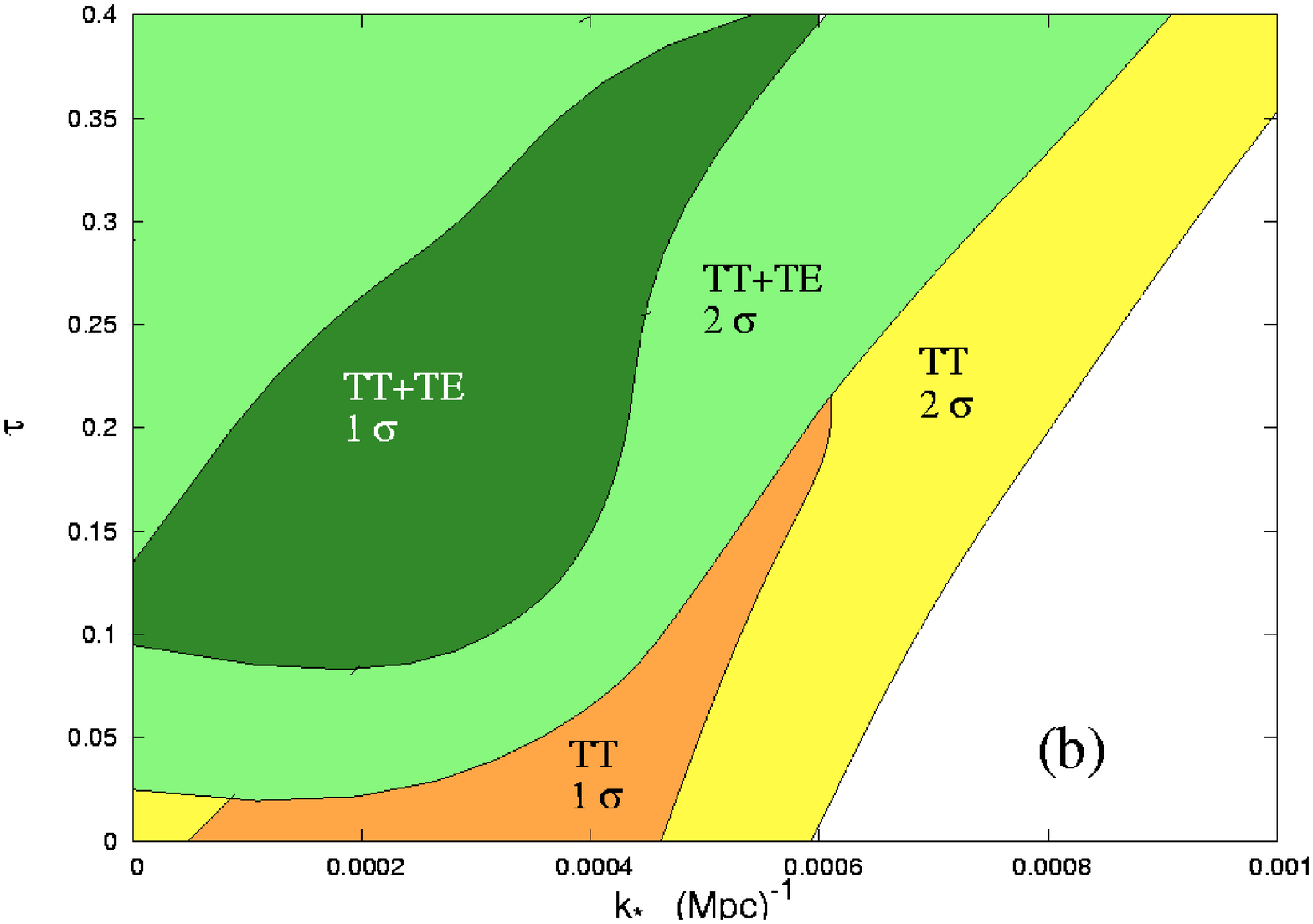}}
\centerline{\epsfxsize=0.6\textwidth\epsfbox{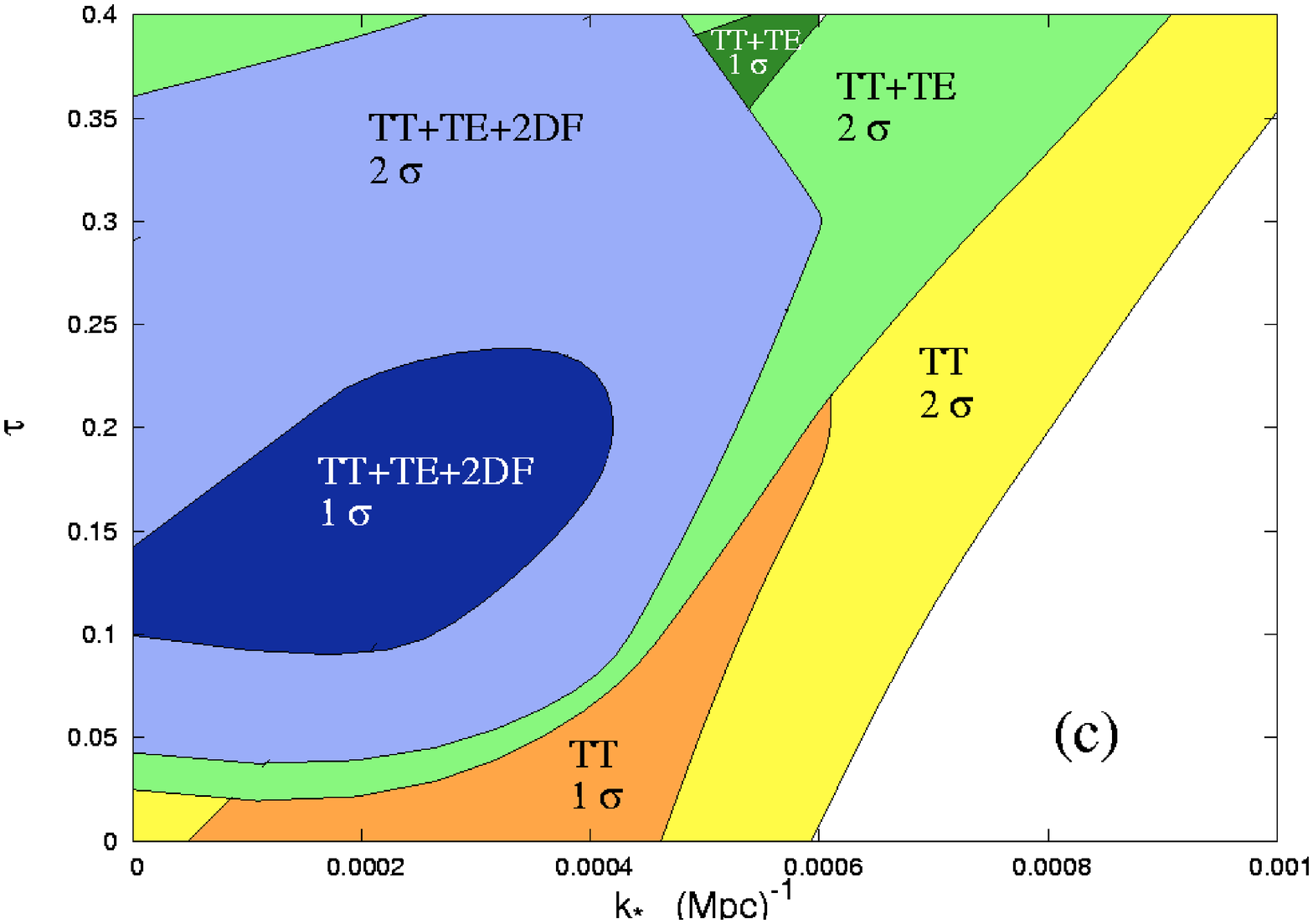}
\epsfxsize=0.63\textwidth\epsfbox{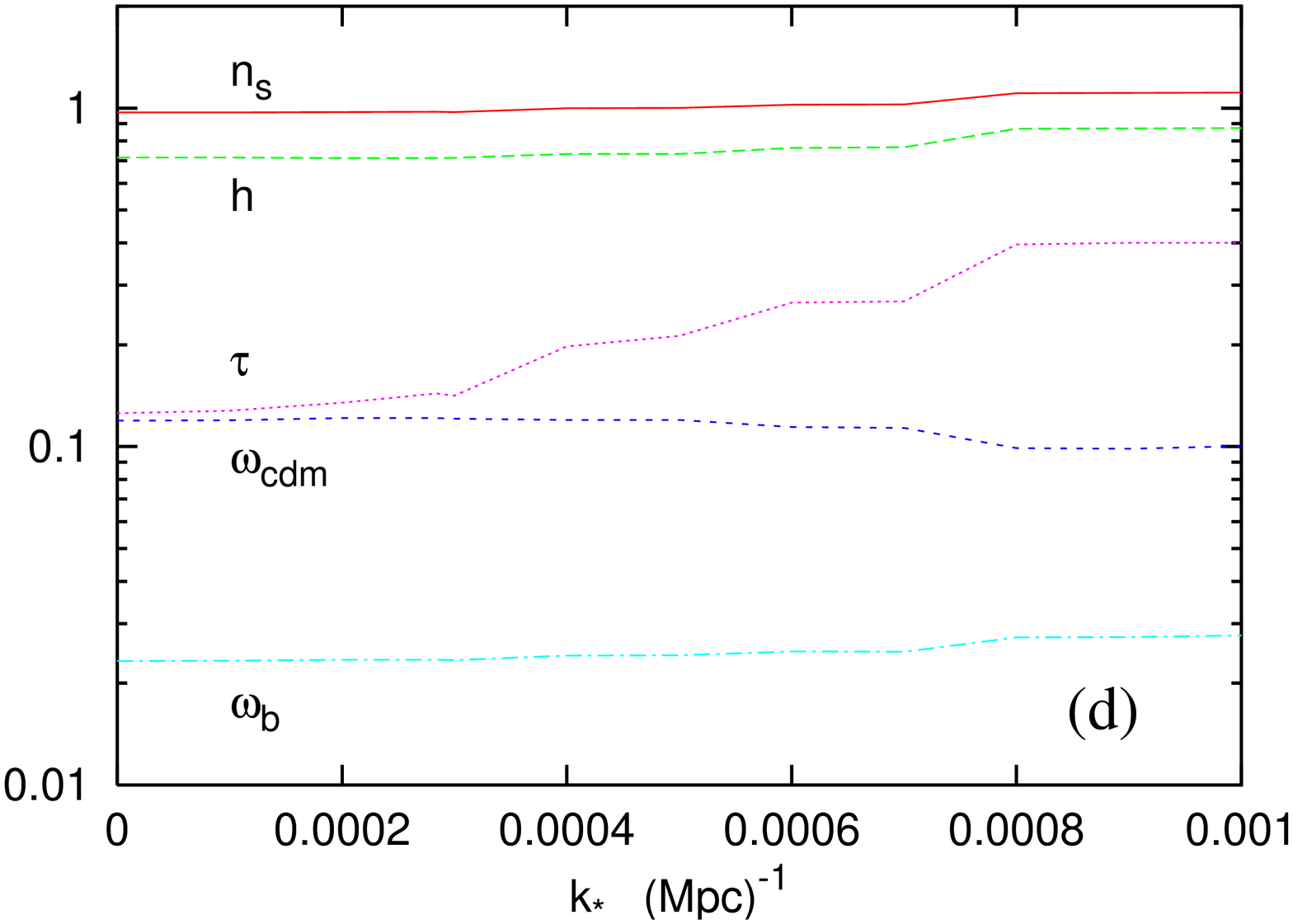}}
\caption{\small (a)-(c): 1 and 2 $\sigma$ confidence regions in the $\tau$-$k_*$
plane for WMAP TT, TT$+$TE and TT$+$TE$+$2dF data sets.  (d) Most probable values
of other parameters as a function of $k_*$, in the case of the TT$+$TE$+$2dF 
data sets.}}

To give a more precise idea of the statistical significance, we tabulate the best fit
parameters and $\chi^2$ of the fit for the three cases in Table 1.  As expected,
the TT data by themselves show the strongest preference for a cutoff, but the
change in $\chi^2$ is still only 2.6.  When the TE data are included $\Delta\chi^2$
falls to 1.8.  With the inclusion of 2dF data, $\Delta\chi^2=2.0$.  Thus the model
with no cutoff is within $1.4 \sigma$ of the minimum $\chi^2$ point.

\TABLE{
\begin{tabular}{|c|c|c|c|c|c|c|c|}
\hline
parameter & WMAP best fit & \multicolumn{2}{c}{WMAP TT} \vline & 
\multicolumn{2}{c}{TT$+$TE} \vline &
\multicolumn{2}{c}{TT$+$TE$+$2dF} \vline\\
\cline{3-8}
 & $\Lambda$CDM model&  $k_*=0$ & $k_*\neq 0$ & $k_*=0$ & $k_*\neq 0$ & $k_*=0$ & $k_*\neq 0$ \\
\hline\hline
 \( k_{*} / 10^{-4} \)Mpc$^{-1}$  & $-$  &  $-$ & 3.4 & $-$ & 2.7 & $-$ & 2.8 \\
\hline
 \( \Omega_b h^2 \) &  0.023& 0.023 & 0.023 & 0.023 & 0.023 & 0.023 & 0.023 \\
\hline
\( \Omega_{m}h^2\) & 0.15 &  0.15 & 0.15 & 0.15 & 0.15 & 0.14 & 0.14 \\
\hline
\( h\)  &  0.68 & 0.69 & 0.69 & 0.69 & 0.70 & 0.71 & 0.71 \\
\hline
\( \tau \) &  0.11 & 0 & 0 & 0.11 & 0.12 &0.13 & 0.14 \\
\hline
\( n_s \) &  0.97 &  0.95 & 0.95 & 0.97 & 0.97 & 0.97 & 0.98 \\
\hline
\( {\chi^2}^{\phantom{1}} \)  &  1431 & 972.2 & 969.6 & 1431.3 &  1429.5 & 1458.7 & 1456.7 \\
\hline 
 d.o.f.  &  1342 & 893 &  892 &  1342 &   1341 &  1374 &  1373 \\
\hline
\end{tabular}
\caption{Maximum likelihood parameters for the three different data sets, with and without
cutoff $k_*$.  WMAP best fit is from table 1 of \cite{map-params}.}
}

The above results were obtained using the ansatz (\ref{ansatz}) for $P(k)$.  We repeated the
analysis for primordial spectra of the forms shown in figs. 2a and 2b, as well as a sharp
step-function cutoff.  They give no better nor worse a fit relative to (\ref{ansatz}).

\section{Summary and conclusions}

We have investigated whether the anomalously small power of large-angle CMB temperature
anisotropies is indicative of new physics that could suppress the primoridal power spectrum
of inflaton fluctuations at low wave number by introducing a cutoff $k_*$.  We analyzed the
complete WMAP data set including polarization, with and without the addition of the 2dF
galaxy power spectrum data.  Allowing for a variation of the six standard cosmological
parameters characterizing a spatially flat universe, we find a marginal preference for a
nonvanishing cutoff scale of $k_* = (2.7\pm 1)\times 10^{-4}$ Mpc$^{-1}$, in agreement with
ref.\ \cite{Bridle}. Like them we find that the likelihood is nongaussian for low $k$ (fig.
5) so that $k_*$ is bounded to be above zero at a confidence level of only $59-64$ \%
(depending upon whether 2dF data is included).  By contrast, ref.\ \cite{CPKL} finds
$k_*>0$  at the $87-93$ \% c.l. Moreover their most likely value of $k_*$ is larger than 
ours,  $(4.9\pm1)\times 10^{-4}$ Mpc$^{-1}$.  We have explained the reasons for the
differences between these results and our own.

Thus we conclude that at present the motivation from the low quadrupole and octopole
moments for a power spectrum with an infrared cut-off is quite weak.  Nevertheless we
have pointed out correlations between the parameters which will be relevant for fitting
the low quadrupole with such models.  In particular there is a tendency for optical
depth $\tau$ to increase in order that the low-$l$ polarization data remain consistent
with the model in the presence of a low-$k$ cutoff.

How can we reconcile the small statistical significance of the need for a cutoff with the
WMAP collaboration's statement that the probability of having such low power at large angles
is only $2\times 10^{-3}$?  The answer lies in the highly nonGaussian nature of the
likelihood function for the $C_l$'s \cite{Verde,BJK}.   Expressing the temperature anisotropy as
$\delta T = \sum_{lm} a_{lm} Y_{lm}(\phi,\theta)$, the $a_{lm}$'s have a Gaussian
distribution, but the multipole moments, $C_l = \sum_m\langle |a_{lm}|^2 \rangle$, have the
distribution \cite{knox}
\beq
\label{prob}
{dP\over dC_l} \ \propto\  x^{l-1/2} e^{-(l+1/2)x}\,;\qquad  x \equiv {C_l^{\rm obs}\over
C_l^{\rm theo}}. 
\eeq
For large $l$ this can be well-approximated by a Gaussian, but for low
$l$ it is quite asymmetric about the average value $x=1$.  From this fact we can
easily reconcile the small probability $2\times 10^{-3}$ with the lack of a significant
improvement in the $\chi^2$ when considering models with cutoff power spectra, 
as we now explain. 

Let us review the method which was used to arrive at WMAP's small probability, as  described
in section 7 of \cite{map-params}.  From a large ensemble of models in the vicinity of the
maximum likelihood model, one generates simulations of the data which include the effects of
cosmic variance and the sky cut which is applied to the actual WMAP observations.  Among
these realizations, one then counts the number whose value of the quadrupole
moment $C_2$ is less than or equal to the measured $C_2^{\rm obs}$, and compares to the
total number in the ensemble.  If we neglect the experimental noise 
(which is much smaller than cosmic
variance at l=2) and the sky cut (which correlates the quadrupole with
other multipoles), then for the quadrupole this amounts to computing
\beq
\label{PC2}
	P[C_2 < C_2^{\rm obs}] = {\int_0^{0.1} dx {dP\over dC_2} \over \int_0^\infty dx
{dP\over dC_2} } \cong 0.008
\eeq
where we used the observed value $C_2^{\rm obs}=123$ and the theoretical one $C_2^{\rm
theo}=1200$. Evidently inclusion of the ``experimental complications'' decreases this
probability somewhat, but at least we can roughly understand its small order of magnitude.

Such an analysis could in principle be repeated including a cutoff power spectrum. We
can roughly anticipate the results using the same approximation as in equation (4.2),
but replacing the $\Lambda$CDM integration bound $x_{\Lambda{\rm CDM}} = 0.1$ by
$x_{k_*} \simeq 0.2$, since the cutoff model is able to reduce the theoretical
quadrupole by about a factor of 2 (further reduction comes at the expense of dragging
down the higher multipoles too much). Then we would obtain $P[C_2 < C_2^{\rm obs}] = 
0.04$. This is a five-fold increase over the low probability in \pref{PC2}, but it is
still low.  Neither model satisfactorily
explains the observed low quadrupole, unless large statistical fluctuations are invoked.
Because of the nongaussianity of the likelihood function, only a much more radical
suppression of $C_l^{\rm theo}$ could significantly increase this probability.

This explains why in our Bayesian analysis, we do not
see a big difference between the best $\chi^2$ value for $k_* \neq 0$ and
that for $k_*=0$. In this approach, we are making a different comparision,
namely:
\beq
	\Delta\chi^2 \sim -2\left( \ln {dP\over dC_2}(x_{\Lambda\rm CDM}) - 
	\ln {dP\over dC_2}(x_{k_*}) \right) \cong -1.6
\eeq
This crude estimate does not replace the  global seven-parameter analysis that we
performed, but it gives a very good approximation for the improvement in $\chi^2$
which we find. From this argument we conclude that there is no real discrepancy between 
WMAP's low probability $2\times 10^{-3}$ and the seemingly much higher probability found
by us and by \cite{GWMMH,Bridle,CPKL}. However we also conclude that both the standard
$\Lambda$CDM model and the low-$k$ cutoff models are rather poor fits to the observed
quadrupole, and to do better one should find a way to more effectively suppress the
theoretical value of the quadrupole moment.    

{\bf Note added:}  After the first version of this work appeared, ref.\ \cite{FZ} proposed a
realization of the kind of cutoff spectrum we have considered here.  Their claim of a
2.5$\sigma$ signal (in version 1 of their paper) is subject to the same observations as we
have made with regard to \cite{CPKL}, since they apply exactly the same analysis.  We also
became aware of ref. \cite{Jing,Yoko}, which considered theoretical implication of the low
quadrupole moment when it was first measured by COBE.

\acknowledgments 
We thank L.\ Verde for her kind assistance with the WMAP likelihood code,
and A.\ Lewis, D.\ Spergel, I.\ Tkachev and L.\ Verde, for valuable discussions concerning
the last point of our conclusions. We also thank A.\ Cooray for interesting comments on the
manuscript, and the referee for useful suggestions.

The research of JC is partially supported by grants from NSERC. (Canada) and NATEQ
(Qu\'ebec).

\end{document}